\documentclass[twocolumn,prd]{revtex4}

\usepackage{graphicx}% Include figure files
\usepackage{dcolumn}% Align table columns on decimal point
\usepackage{bm}% bold math

\newcommand{\be}{\begin{equation}}
\newcommand{\ee}{\end{equation}}
\newcommand{\ba}{\begin{eqnarray}}
\newcommand{\ea}{\end{eqnarray}}

\begin{document}
\title{Ultra-High Energy Neutrino Fluxes and Their Constraints}
\author{Oleg~E.~Kalashev$^a$, Vadim~A.~Kuzmin$^a$, Dmitry~V.~Semikoz$^{a,b}$,
G{\"u}nter Sigl$^c$}
\affiliation{$^a$ Institute for Nuclear Research of the Academy
of Sciences of Russia,\\
Moscow, 117312, Russia\\
$^b$ Max-Planck-Institut f\"ur Physik (Werner-Heisenberg-Institut),\\
F\"ohringer Ring 6, 80805 M\"unchen, Germany\\
$^c$ GReCO, Institut d'Astrophysique de Paris, C.N.R.S., 98 bis boulevard
Arago, F-75014 Paris, France}

\begin{abstract}
Applying our recently developed propagation code we review extragalactic
neutrino fluxes above $10^{14}\,$eV in various scenarios
and how they are constrained by current data. We specifically identify
scenarios in which the cosmogenic neutrino flux above
$\simeq10^{18}\,$eV, produced by pion
production of ultra high energy cosmic rays outside their
sources, is considerably higher than the "Waxman-Bahcall bound".
This is easy to achieve for sources with hard injection spectra and
luminosities that were higher in the past.
Such fluxes would significantly increase the chances to detect
ultra-high energy neutrinos with experiments currently under
construction or in the proposal stage.
\end{abstract}

\maketitle

\section{Introduction}
The farthest source so far observed in neutrinos was supernova
SN1987A from which about 20 neutrinos in the 10-40 MeV
range where detected~\cite{sn1987A}. Extraterrestrial neutrinos
of much higher energy are usually expected to be produced
as secondaries of cosmic rays interacting with ambient matter
and photon fields and should thus be associated with cosmic
ray sources ranging from our Galaxy to powerful active
galactic nuclei (AGN)~\cite{review_nu}. Traditional neutrino telescopes
now under construction aim to detect such neutrinos up to
$\sim10^{16}\,$eV by looking for showers and/or tracks
from charged leptons produced by charged current reactions of
neutrinos in ice, in case of AMANDA~\cite{amanda,amanda_limit}
and its next generation version ICECUBE~\cite{icecube},
in water, in case of BAIKAL \cite{baikal,baikal_limit}, 
ANTARES~\cite{antares}, and NESTOR ~\cite{nestor}, or underground,
in case of MACRO ~\cite{MACRO} (for recent reviews of neutrino
telescopes see Ref.~\cite{nu_tele}).

On the other hand, the problem of the origin of the cosmic rays
themselves is still unsolved, especially at ultra-high energies (UHE)
above $\simeq4\times10^{19}\,$eV, where they lose energy rapidly
by pion production and pair production (protons only) on the cosmic
microwave background (CMB)~\cite{reviews,bs}. For sources further away
than a few dozen
Mpc this would predict a break in the cosmic ray flux known as
Greisen-Zatsepin-Kuzmin (GZK) cutoff~\cite{gzk}, around
$50\,$EeV. This break has not been observed by experiments such as
Fly's Eye~\cite{Fly}, Haverah Park~\cite{Haverah},
Yakutsk~\cite{Yakutsk} and AGASA~\cite{agasa}, which
instead show an extension beyond the expected GZK cutoff and events
above $100\,$EeV (however, the new experiment HiRes~\cite{Hires}
currently seems to see a cutoff in the monocular data~\cite{Hires2001}).
This has lead to the current construction of the southern site of
the Pierre Auger Observatory~\cite{auger}, a combination of an array of charged
particle detectors with fluorescence telescopes for air showers
produced by cosmic rays above $\sim10^{19}\,$eV which will
lead to an about hundred-fold increase of data. The telescope array,
another planned project based on the fluorescence technique,
may serve as the optical component of the northern Pierre Auger site
planned for the future~\cite{ta}. There are also plans
for space based observatories such as EUSO~\cite{euso} and OWL~\cite{owl}
of even bigger acceptance. These instruments will also have
considerable sensitivity to neutrinos above $\sim10^{19}\,$eV,
typically from the near-horizontal air-showers that are produced
by them~\cite{auger_nu}. Furthermore, the old Fly's Eye
experiment~\cite{baltrusaitis} and the AGASA experiment~\cite{agasa_nu}
have established upper limits on neutrino fluxes based on the
non-observation of horizontal air showers.

In addition, there are plans to construct telescopes to detect
fluorescence and \v{C}erenkov light from near-horizontal showers produced in
mountain targets by neutrinos in the intermediate window of
energies between $\sim10^{15}\,$eV and $\sim10^{19}\,$eV~\cite{fargion,mount}.
The alternative of detecting neutrinos by triggering onto the
radio pulses from neutrino-induced air showers is also investigated
currently~\cite{radhep}. Two implementations of this technique,
RICE, a small array of radio antennas in the South pole ice~\cite{rice},
and the Goldstone Lunar Ultra-high energy neutrino Experiment (GLUE),
based on monitoring of the moons rim with the NASA Goldstone
radio telescope for radio pulses from neutrino-induced
showers~\cite{glue}, have so far produced neutrino flux upper limits.
Acoustic detection of neutrino induced interactions is also
being considered~\cite{acoustic}.

The neutrino detection rates for all future instruments will crucially
depend on the fluxes expected in various scenarios. The flux of "cosmogenic"
neutrinos created by primary protons above the GZK cutoff in interactions
with CMB photons depends both on primary proton spectrum and on the
location of the sources. The cosmogenic neutrino flux is the only
one that is guaranteed to exist just by the observations of
ultra-high energy cosmic rays (UHECRs)
and was studied soon after the discovery of the CMB~\cite{cosmogenic}.
Note, however, that there is no firm lower bound
on the cosmogenic neutrino flux if the UHECR
sources are much closer than the GZK distance.

If sources are located beyond the GZK distance and the proton flux
extends beyond the GZK cutoff, the neutrino fluxes
can be significant. This possibility is favored by the lack
of nearby sources and by the hardening of the cosmic ray spectrum
above the ``ankle'' at $\simeq5\times10^{18}\,$eV. It is also
suggested by possible  correlations of UHECR arrival directions with
compact radio quasars~\cite{corr_radio} and more significant correlations
with  BL Lacertae objects~\cite{corr_bllac},
some of which possibly also luminous in GeV $\gamma-$rays~\cite{egret_bllac}
and at distances too large to be consistent with the absence of the GZK cutoff.

Whereas roughly homogeneously distributed proton sources can naturally
explain the UHECR flux below the GZK cutoff (see, e.g.,
Refs.~\cite{berezinsky2002} and~\cite{photons}),
the highest energy events may represent a new component. They
may be new messenger particles, which propagate through the Universe
without interacting with the CMB~\cite{messengers}, or may have
originated as extremely high energy ($E\gtrsim10^{23}\,$eV)
photons, which can propagate several hundred Mpc (constantly loosing energy)
and can create secondary photons inside the GZK volume~\cite{photons}.
Decaying super-heavy relics from the early
Universe (see Ref.~\cite{bs,kt} for reviews) can also explain UHECRs
and predict UHE neutrino fluxes detectable by future experiments. 

Another speculative possibility is to explain UHECRs beyond the GZK
cutoff by the UHE protons and photons from decaying Z-bosons produced
by UHE neutrinos interacting with the relic neutrino
background~\cite{zburst1}.  
The big drawback of this scenario is the need of enormous
primary neutrino fluxes that cannot be produced by known astrophysical
acceleration sources without overproducing the GeV
photon background~\cite{kkss}, and thus most likely
requires a more exotic top-down type source such as X particles
exclusively decaying into neutrinos~\cite{gk}. As will be shown
in Sect.~VI, even this possibility is also significantly 
constrained by existing measurements.

Active galactic nuclei (AGN) can be sources of neutrinos if protons 
are accelerated in them~\cite{review_nu}. In the present paper
we consider only the two representative limits of low and high
optical depth for pion (and neutrino) production in the source.
In the first case the protons accelerated in the AGN freely escape
and neutrinos are produced only in interactions with the CMB
(cosmogenic neutrinos). For the second case we discuss an
example of possible high neutrino
fluxes from a non-shock acceleration AGN model~\cite{neronov},
in which primary protons lose all their energy and produce neutrinos
directly in the AGN core.

Motivated by the increased experimental
prospects for ultra-high energy neutrino detection, in the present
paper we reconsider flux predictions in the above scenarios
with our recently combined propagation codes~\cite{code1,code2,photons,kkss}.
Our main emphasize is thereby on model independent flux ranges
consistent with all present data on cosmic and $\gamma-$rays.
For any scenario involving pion production the fluences of the
latter are comparable to the neutrino fluences.
However, electromagnetic
 (EM) energy injected above $\sim10^{15}\,$eV cascades down to below the
 pair production threshold for photons on the CMB  and EM energy above 100 GeV also
 cascades down
 due to the pair production on infrared/optical background. 
 The resulting intensity and spectrum of
 $\gamma-$rays below 100 GeV is rather insensitive to
 these backgrounds~\cite{berez75,ca} (for review see \cite{bs}). 
The cascade
thus gives rise to a
diffuse photon flux in the GeV range which is constrained by the flux observed
by the EGRET instrument on board the Compton $\gamma-$ray
observatory~\cite{egret}. For all neutrino flux scenarios the
related $\gamma-$ray and cosmic ray fluxes have to be consistent
with the EGRET and cosmic ray data, respectively.

Sect.~II summarizes the numerical technique used in this paper.
In Sect.~III we discuss the cosmogenic neutrino flux and its
dependence on various source characteristics. We specifically
find an upper limit considerably higher than typical fluxes
in the literature and remark why it is higher than the
Waxman-Bahcall (WB)~\cite{wb} and even the
Mannheim-Protheroe-Rachen (MPR)~\cite{mpr}
bounds for sources transparent to cosmic and $\gamma-$rays.
In Sect.~IV we review neutrino flux predictions in top down
scenarios where UHECRs are produced in decays of
super-massive particles continuously released from topological
defect relics from the early Universe.
Sect.~V discusses neutrino fluxes in scenarios where the
cosmic rays observed at the highest energies are produced as
secondaries of these neutrinos from interactions with the
relic cosmological neutrino background, often called Z-burst
scenario. In Sect.~VI we focus on a combination of top-down
and Z-burst scenarios considered by Gelmini and Kusenko~\cite{gk},
namely super-heavy particles mono-energetically decaying exclusively
into neutrinos"(see, however, \cite{berkaros}). In Sect.~VII  we discuss possible high neutrino
fluxes from a non-shock acceleration AGN model~\cite{neronov}.
Finally, in Sect.~VIII we conclude.

\section{Numerical Technique}
Our simulations are based on two independent codes that
have extensively been compared down to the level of
individual interactions. Both of them are implicit transport codes
that evolve the spectra of nucleons, $\gamma-$rays, electrons,
electron-, muon-, and tau-neutrinos, and their antiparticles
along straight lines. Arbitrary injection spectra and
redshift distributions can be specified for the sources and
all relevant strong, electromagnetic, and weak interactions
have been implemented. For details see Refs.~\cite{code1,code2,kkss}.

Relevant neutrino interactions for the Z-burst scenario
are both the s-channel production of
Z bosons and the t-channel production of W bosons.
The decay products of the Z boson were taken from simulations
with the ~\cite{PYTHIA} Monte Carlo event generator using the
tuned parameter set of the OPAL Collaboration~\cite{OPAL}.

The main ambiguities in propagation of photons concern the unknown
rms magnetic field strength $B$ which can influence the
predicted $\gamma-$ray spectra via synchrotron cooling
of the electrons in the EM cascade, and the strength of the
universal radio background (URB) which influences pair production
by UHE $\gamma-$rays~\cite{pb}. Photon interactions in the GeV to
TeV range are dominated by infrared and optical universal photon
backgrounds (IR/O), for which we took the results of Ref. \cite{Primack}.
The resulting photon flux in GeV range in not sensitive to
details of the IR/O backgrounds. 

Concerning the cosmology parameters we chose the Hubble parameter
$H_0 = 70~ {\rm km}~{\rm s}^{-1}~ {\rm Mpc}^{-1}$ and a
cosmological constant $\Omega_\Lambda = 0.7$, as favored today.
These values will be used in all cases unless otherwise indicated.

For the neutrinos we assume for simplicity that all  three flavors  are
completely mixed as suggested by experiments~\cite{superK}
and thus have equal fluxes. For each flavor we sum fluxes of  
particles and antiparticles.

Predictions for the all fluxes in both codes agree within tens of percents.
Only for the Z-burst scenarios the photon fluxes agree only
within a factor $\simeq2$ between
the two original codes due to the different implementation of
Z-decay spectra and the ambiguities in photon propagation
mentioned above. This difference has no influence on the conclusions
of this paper.

In the present investigation we parameterize power law
injection spectra of either protons (for UHECR sources) or
neutrinos (for Z-burst models) per co-moving volume
in the following way:
\begin{eqnarray}
  \phi(E,z)&=&f(1+z)^m\,E^{-\alpha}\Theta(E_{\rm max}-E)\,
  \nonumber\\
  &&z_{\rm min}\leq z\leq z_{\rm max}\,,\label{para_inj}
\end{eqnarray}
where $f$ is the normalization that has to be fitted to the
data. The free parameters are the spectral index $\alpha$, the maximal
energy $E_{\rm max}$, the minimal and maximal
redshifts $z_{\rm min}$, $z_{\rm max}$, and the redshift
evolution index $m$. The resulting neutrino spectra depend insignificantly on
$z_{\rm min}$ in the range $0\leq z_{\rm min}\lesssim0.1$
where local effects could play a role, and thus we will set
$z_{\rm min}=0$ in the following.

To obtain the maximal neutrino fluxes for a given set of values for
all these parameters , we determine the maximal normalization $f$
in Eq.~(\ref{para_inj}) by demanding that both the accompanying
nucleon and $\gamma-$ray
fluxes are below the observed cosmic ray spectrum and
the diffuse $\gamma-$ray background observed by EGRET,
respectively.

\section{The Cosmogenic Neutrino Flux}

\subsection{Dependence on unknown parameters}

\begin{figure}[ht]
\includegraphics[height=0.5\textwidth,clip=true,angle=270]{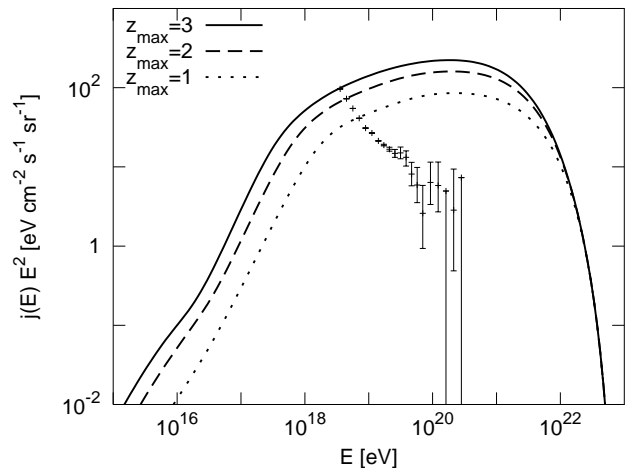}
\caption[...]{Dependence of the average maximal cosmogenic neutrino
flux per flavor consistent with all cosmic and $\gamma-$ray data on the
maximal redshift $z_{\rm max}$, for the values indicated.
Values assumed for the other parameters in Eq.~(\ref{para_inj}) for
proton primaries are $E_{\rm max}=10^{23}\,$eV, $m=3$, $\alpha=1.5$.
Also shown are the AGASA cosmic ray data above
$3\times10^{18}\,$eV~\cite{agasa}.}
\label{F1}
\end{figure}

\begin{figure}[ht]
\includegraphics[height=0.5\textwidth,clip=true,angle=270]{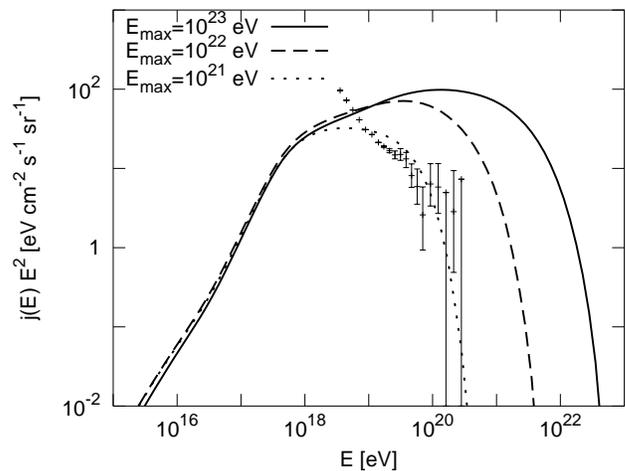}
\caption[...]{Dependence of the average maximal cosmogenic neutrino
flux per flavor consistent with all cosmic and $\gamma-$ray data on the
maximal injection energy $E_{\rm max}$, for the
values indicated. Values assumed for the other parameters are
$z_{\rm max}=2$, $\alpha=1.5$, $m=3$. The cosmic ray
data are as in Fig.~\ref{F1}.}
\label{F2}
\end{figure}

\begin{figure}[ht]
\includegraphics[height=0.5\textwidth,clip=true,angle=270]{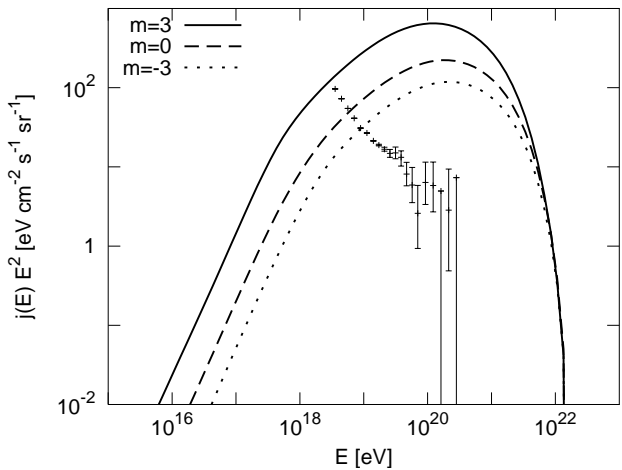}
\caption[...]{Dependence of the average maximal cosmogenic neutrino
flux per flavor consistent with all cosmic and $\gamma-$ray data on the
source evolution index $m$, for the
values indicated. Values assumed for the other parameters are
$z_{\rm max}=2$, $E_{\rm max}=3\times10^{22}\,$eV,
$\alpha=1$. The cosmic ray data are as in Fig.~\ref{F1}.}
\label{F3}
\end{figure}

\begin{figure}[ht]
\includegraphics[height=0.5\textwidth,clip=true,angle=270]{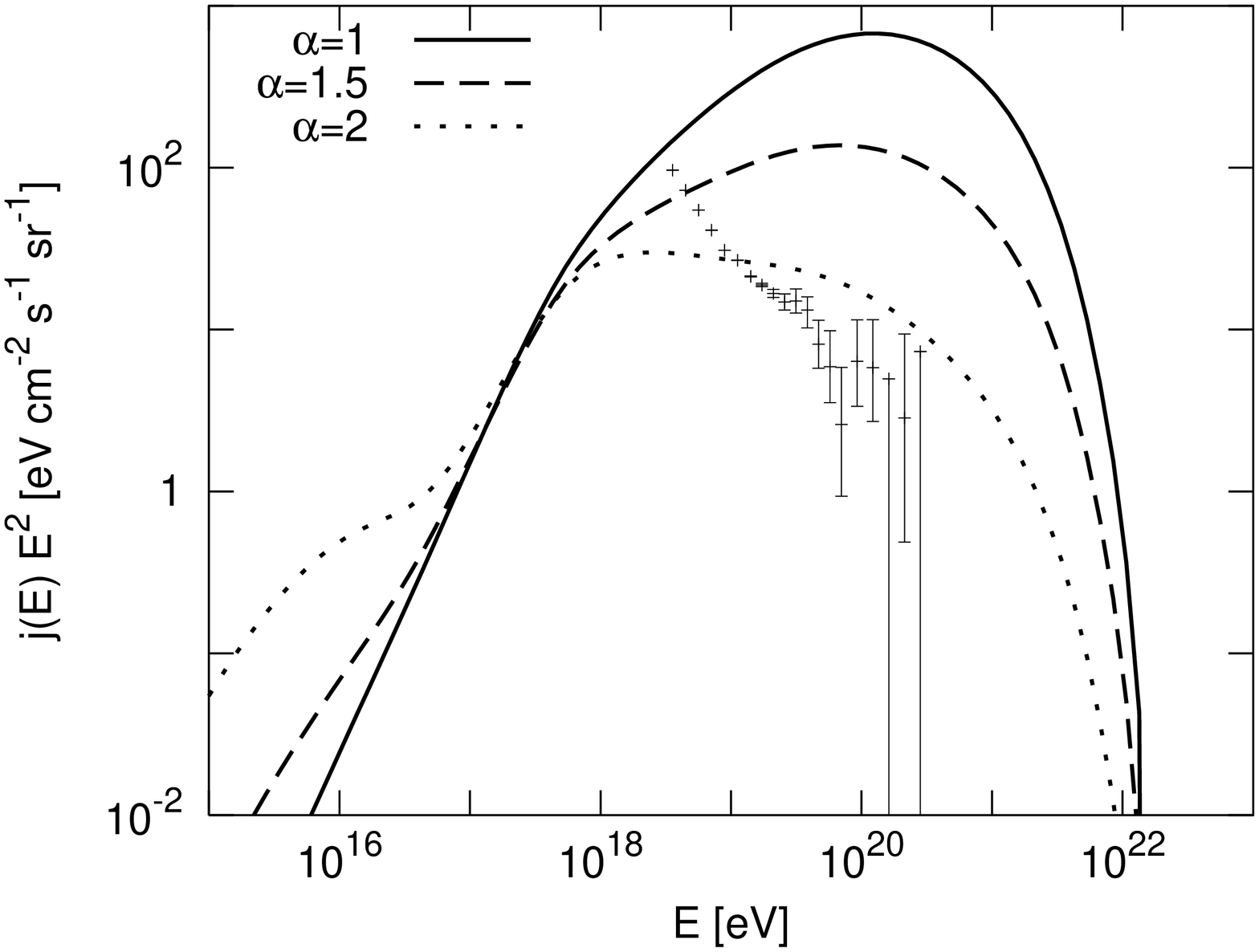}
\caption[...]{Dependence of the average maximal cosmogenic neutrino
flux per flavor consistent with all cosmic and $\gamma-$ray data on the
injection spectrum power law index $\alpha$, for the
values indicated. Values assumed for the other parameters are
$z_{\rm max}=2$, $E_{\rm max}=3\times10^{22}\,$eV,
$m=3$. The cosmic ray data are as in Fig.~\ref{F1}.}
\label{F4}
\end{figure}

\begin{figure}[ht]
\includegraphics[height=0.5\textwidth,clip=true,angle=270]{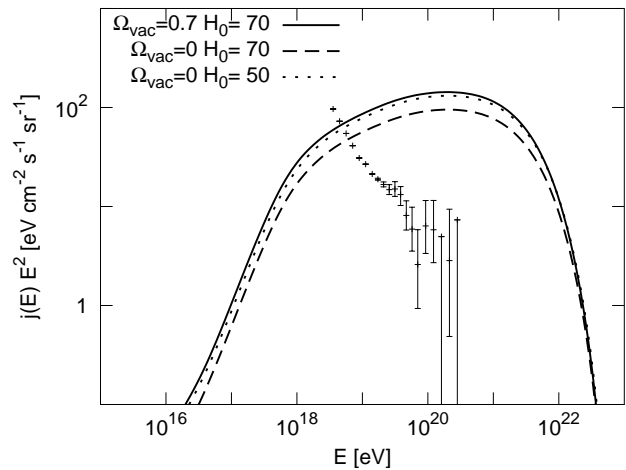}
\caption[...]{Dependence of the average maximal cosmogenic neutrino
flux per flavor consistent with all cosmic and $\gamma-$ray data on the
cosmological vacuum energy density $\Omega_\Lambda$ and Hubble rate
$H_0$, measured in ${\rm km}{\rm s}^{-1}{\rm Mpc}^{-1}$, for the values
indicated. Values assumed for the other parameters are
$z_{\rm max}=2$, $E_{\rm max}= 10^{23}\,$eV, $\alpha=1.5$,
$m=3$. The cosmic ray data are as in Fig.~\ref{F1}.}
\label{F4c}
\end{figure}

In this section we discuss the case when  primary UHECRs produce 
cosmogenic neutrinos as well as $\gamma-$rays during propagation.
For EM propagation we
use $B=10^{-9}$ G and the intermediate URB strength estimate of
Ref.~\cite{pb}. These parameters only influence the $\gamma-$ray
flux at UHEs, but not in the GeV range where the flux only depends
on the total injected EM energy. Therefore, in this scenario the
resulting neutrino fluxes are insensitive to the poorly known UHE
$\gamma-$ray absorption because the ``visible'' UHE flux is always
dominated by the primary cosmic rays and not by the secondary
$\gamma-$ray flux, as can be seen in Figs.~\ref{F5}
and~\ref{F5a} below.

Figs.~\ref{F1}, \ref{F2}, \ref{F3}, \ref{F4}, and~\ref{F4c} show the
dependencies of the  cosmogenic neutrino flux (average
per flavor) for which associated cosmic and $\gamma-$ray fluxes
are consistent with the data, on the maximal source redshift
$z_{\rm max}$, maximal injection energy $E_{\rm max}$, redshift
evolution index $m$, spectral power law index $\alpha$, and
cosmological parameters, respectively.  In each figure the line for
the highest neutrino flux corresponds to a significant contribution
of the accompanying $\gamma$-ray flux to the observed flux at the EGRET
region, whereas for the other lines the $\gamma$-ray flux gives negligible 
contributions to the EGRET flux. 

Fig.~\ref{F2} shows that a change of the primary proton maximum energy
changes the secondary neutrino flux only in the high energy region.
The reason for this behavior is the shape of the pion production
cross section which at the lowest energies is dominated by the single
pion $\Delta-$resonance. Figs.~\ref{F1} and~\ref{F3} show that the
cosmogenic neutrino flux at the lowest energies mostly depends on
the maximum redshift $z_{\rm max}$ and the evolution index $m$.
This is especially relevant for experiments with their main sensitivity
below $\sim10^{19}\,$eV such as ICECUBE (see Fig.~\ref{F5a} below).
Fig. ~\ref{F4} shows that the maximal neutrino flux significantly
increases with decreasing proton injection power law index.
Fig. ~\ref{F4c} shows that the variation of the maximal neutrino
flux with cosmology parameters $\Omega_\Lambda$ and $H_0$ is rather
modest, about 50\% for values discussed in recent years.
   
\subsection{Active galactic nuclei as UHECR sources}

\begin{figure}[ht]
\includegraphics[height=0.5\textwidth,clip=true,angle=270]{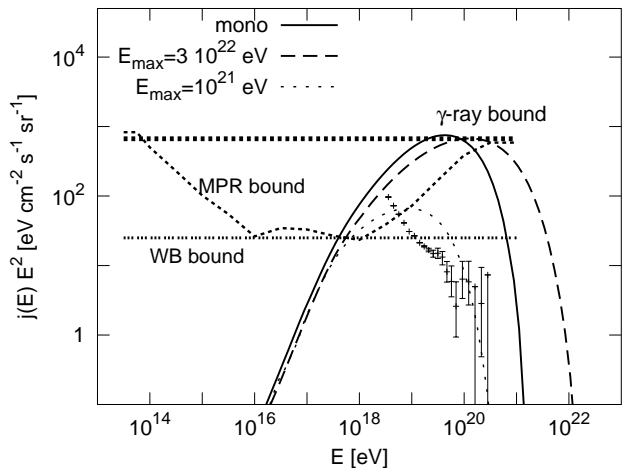}
\caption[...]{Dependence of the average cosmogenic neutrino flux
per flavor on maximum injection energy $E_{\rm max}$, for the values
indicated, assuming $\alpha=1$ and the AGN evolution parameters discussed
in the text. ``mono'' indicates mono-energetic proton injection
at $E=3\times10^{21}\,$eV. For comparison, the $\gamma-$ray bound
derived from the EGRET
GeV $\gamma-$ray flux~\cite{egret}, Eq.(\ref{egret}), the MPR limit
for optically thin  sources~\cite{mpr}, and the WB limit for
AGN-like redshift evolution~\cite{wb}
are also shown. The cosmic ray data are as in Fig.~\ref{F1}.}
\label{F4a}
\end{figure}

Here we consider AGN sources for the primary UHECR flux, with the
typical evolution parameters $m=3.4$ for $z<1.9$ and $m=0$ for
$1.9 < z < 2.9$~\cite{agn_evol}. The only free remaining parameters
are the power law index $\alpha$ and the maximum energy $E_{\rm max}$ for
the proton injection spectrum, and the flux normalization $f$ in
Eq.~(\ref{para_inj}). We first checked that for the case $\alpha=2$
we agree with the WB bound. Our cosmogenic neutrino fluxes agree
reasonably well with the corresponding ones shown in Fig.~4 of
Ref.~\cite{ess} when taking into account maximal mixing assumed
in the present paper.

Fig.~\ref{F4a} shows the maximal cosmogenic neutrino flux for
sources with hard spectra, $\alpha=1$, as a function of the maximal
proton energy $E_{\rm max}$ and a mono-energetic injection
spectrum at $E=3\times10^{21}\,$eV.
For the case $\alpha=1$, MPR~\cite{mpr} computed the sum of the
cosmogenic neutrino flux and the neutrino flux directly emitted
by the sources which are assumed to be transparent to neutrons.
Our fluxes shown here only include the cosmogenic flux. Nevertheless
our fluxes overshoot the comparable MPR fluxes by up to a factor
5 at energies below the peak flux. This is most likely due to
a combination of the different cosmology (see Fig.~\ref{F4c})
and a different implementation of multi-pion production which
influences interactions of nucleons at high energies, thus at
high redshift and in turn the low energy tail of cosmogenic neutrinos.

Fig.~\ref{F4a} and also Figs.~\ref{F2}-\ref{F4} demonstrate
in a general way that it is easy to exceed the WB bound and even the MPR
bound for injection spectra harder than about $E^{-2}$.
This is because Waxman-Bahcall restricted themselves
to nucleon injection spectra softer than $E^{-2}$ and sources
smaller than nucleon interaction lengths~\cite{wb}. Thus, their bound does
not directly apply to the cosmogenic neutrino flux. In addition, in
our opinion, their assumptions on the injection spectra are too
narrow: Possible
scenarios with hard injection spectra and the AGN redshift
evolution assumed here (which is the same as the one used
by Waxman-Bahcall) include cases where UHECRs are accelerated
by the electromotive force produced by magnetic fields threading the horizons
of spinning super-massive black holes in the centers of
galaxies~\cite{bg} or by reconnection events around forming
galaxies~\cite{colgate}.

In any scenario involving pion production for the creation
of $\gamma-$rays and neutrinos, the fluxes per flavor are approximately
related by $ F_\nu(E) \approx F_\gamma(E)/3$. Assuming smooth
spectra and comparing with the EGRET $\gamma-$ray fluence,
energy conservation implies
\begin{equation} 
  E^2 F_\nu(E) \lesssim 6\times10^2\,{\rm eV}{\rm cm}^{-2}{\rm s}^{-1}
  {\rm sr}^{-1}\,.\label{egret}
\end{equation}  
This ultimate bound is also shown in Fig.~\ref{F4a} and Fig.~\ref{F4b}
below. It corresponds to the MPR limit for optically thick sources.
The maximal $E^2j(E)$ of the fluxes in Figs.~\ref{F4a} and~\ref{F4b} below
indeed reach this $\gamma-$ray bound Eq.~(\ref{egret}).

Note that the two theoretical bounds shown
in Fig.~\ref{F4a} represent fluxes per neutrino flavor under
the assumption of maximal neutrino mixing. They are thus about a factor
2 lower than in Refs.~\cite{wb,mpr} which show muon neutrino fluxes
in the absence of mixing where the tau-neutrinos fluxes are
negligible and electron neutrino fluxes are about a factor 2
smaller. The WB and MPR bounds represent upper neutrino flux
limits for compact sources such as AGN and $\gamma-$ray bursts in
case of small optical depth for nucleons. We discuss a specific
non-shock acceleration scenario in Sect.~VII, for which both of
the above bounds are not valid because the source optical depth for
nucleons is large (for reviews on AGN neutrino fluxes see, e.g.,
Ref.~\cite{review_nu}). 

\subsection{General case of arbitrary source evolution}

\begin{figure}[ht]
\includegraphics[height=0.5\textwidth,clip=true,angle=270]{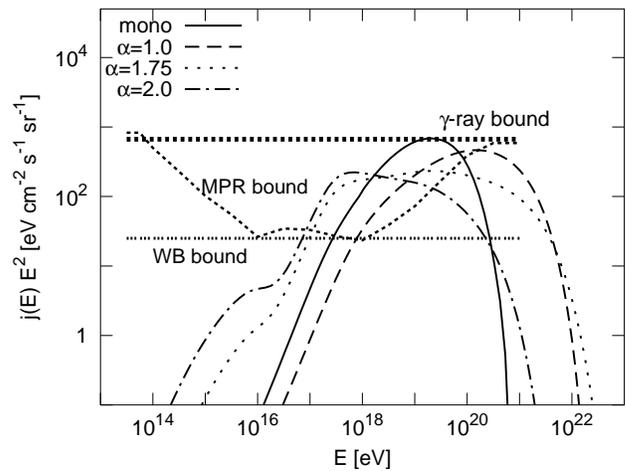}
\caption[...]{Dependence of the average cosmogenic neutrino flux
per flavor maximized over maximal injection energy $E_{\rm max}$,
evolution index $m$, and normalization consistent with all cosmic and
$\gamma-$ray data, on the injection spectrum power law index $\alpha$.
``mono'' indicates mono-energetic proton injection at $E=10^{21}\,$eV.
The rest of the line key is as in Fig.~\ref{F4a}.}
\label{F4b}
\end{figure}

In this section we consider more general UHECR sources and 
relax the restrictions on their redshift evolution. Fig.~\ref{F4b}
shows that cosmogenic neutrino fluxes higher than both the WB
and MPR limits are possible even for relatively soft $E^{-2}$
proton injection spectra, if the redshift evolution is stronger
than for AGNs: The curve for $E^{-2}$ in Fig.~\ref{F4b} corresponds
to the evolution parameters $m=5$, $z_{\rm max}=3$ and
$E_{\rm max}=10^{22}\,$eV, the curve for
$E^{-1.75}$ to $m=4.5$, $z_{\rm max}=3$ and $E_{\rm max}=10^{23}\,$eV,
and the curve for mono-energetic injection to $m=4$, $z_{\rm max}=3$,
and $E_{\rm max}=10^{21}\,$eV.

\begin{figure}[ht]
\includegraphics[width=0.48\textwidth,clip=true]{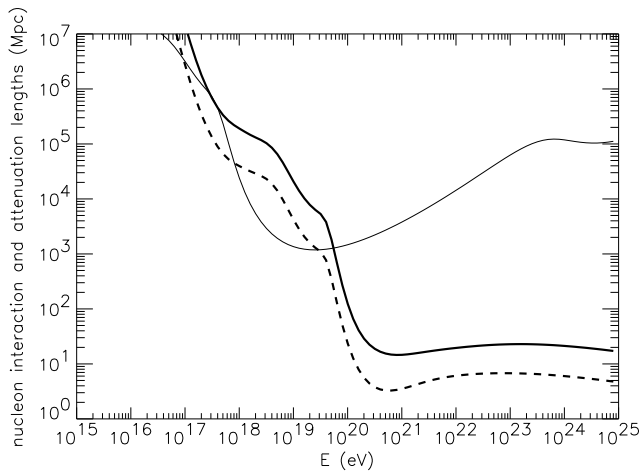}
\caption[...]{The nucleon interaction length (dashed line)
and energy attenuation length (solid line) for photo-pion production
and the proton attenuation length for pair production (thin solid
line) in the combined CMB and the estimated total
extragalactic radio background intensity.}
\label{F6}
\end{figure}

Fig.~\ref{F6} shows that between $\sim10^{18}\,$eV and $\sim10^{20}\,$eV
the energy loss rate of protons on the CMB is dominated by pair
production instead of pion production. The former does not
contribute to neutrino production but the EM cascades initiated
by the pairs lead to contributions to the diffuse $\gamma-$ray background
in the GeV range. Thus, the cosmogenic neutrino flux is the more
severely constrained the bigger the fraction of cosmic ray power
is in the range $10^{18}\,{\rm eV}\lesssim E\lesssim10^{20}\,$eV. This is
mostly important for soft injection spectra and explains why
the total neutrino energy fluence decreases with increasing $\alpha$
in Fig.~\ref{F4b}.

\subsection{Comparison with experimental limits and future sensitivities}

\begin{figure}[ht]
\includegraphics[height=0.5\textwidth,clip=true,angle=270]{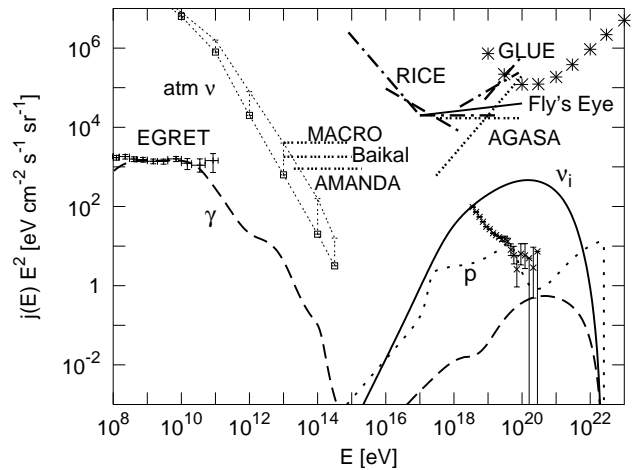}
\caption[...]{A scenario with maximal cosmogenic neutrino fluxes
as obtained by tuning the parameters to $z_{\rm max}=2$,
$E_{\rm max}=10^{23}\,$eV, $m=3$, $\alpha=1$. Also shown are
predicted and observed cosmic ray and $\gamma-$ray fluxes, the
atmospheric neutrino flux~\cite{lipari}, as well as existing upper
limits on the diffuse neutrino fluxes from MACRO~\cite{MACRO},
AMANDA~\cite{amanda_limit},
BAIKAL~\cite{baikal_limit}, AGASA~\cite{agasa_nu}, the
Fly's Eye~\cite{baltrusaitis} and RICE~\cite{rice} experiments, and
the limit obtained with the Goldstone radio telescope (GLUE)~\cite{glue},
as indicated. The cosmic ray data are as in Fig.~\ref{F1}.}
\label{F5}
\end{figure}

\begin{figure}[ht]
\includegraphics[height=0.5\textwidth,clip=true,angle=270]{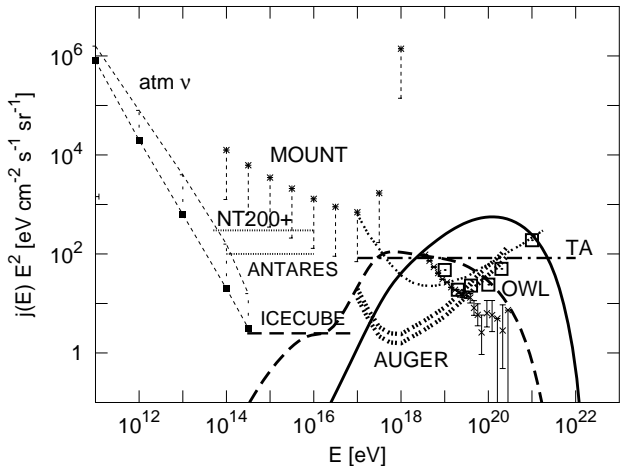}
\caption[...]{The cosmogenic neutrino flux (solid line) shown in Fig.~\ref{F5}
in comparison with expected sensitivities of the currently being constructed
Auger project to electron/muon and tau-neutrinos~\cite{auger_nu},
and the planned projects telescope array (TA)~\cite{ta_nu} (dashed-dotted line), the
fluorescence/\v{C}erenkov detector MOUNT~\cite{mount}, and, indicated by squares, the
space based OWL~\cite{owl_nu} (we take the latter as representative
also for EUSO), the water-based NT200+~\cite{baikal_limit},
ANTARES~\cite{antares} (the NESTOR~\cite{nestor} sensitivity would
be similar to ANTARES according to Ref.~\cite{nu_tele}), and
the ice-based ICECUBE~\cite{icecube}, as indicated. Also shown (dashed line) is
an extreme scenario with $z_{\rm max}=3$,
$E_{\rm max}=10^{22}\,$eV, $m=5$, and $\alpha=2$, leading to a
cosmogenic neutrino flux extending to relatively low energies
where ANTARES and ICECUBE will be sensitive, and the atmospheric
neutrino flux for comparison.}
\label{F5a}
\end{figure}

Figs.~\ref{F5} and~\ref{F5a} shows a scenario maximized over all 4
parameters in comparison to existing neutrino flux upper limits
and expected sensitivities of future projects, respectively. Both
of these fall into two groups, detection in water or ice or underground,
typically sensitive below $\simeq10^{16}\,$eV, and air shower
detection, usually sensitive at higher energies. Existing upper
limits come from the underground MACRO experiment~\cite{MACRO}
at Gran Sasso, AMANDA~\cite{amanda_limit} in the South Pole ice,
and the Lake BAIKAL neutrino telescope~\cite{baikal_limit} in the
first category, and the AGASA ground array~\cite{agasa_nu}, the
former fluorescence experiment Fly's Eye~\cite{baltrusaitis}, the
Radio Ice \v{C}erenkov Experiment RICE~\cite{rice}, and the Goldstone
Lunar Ultra-high energy neutrino experiment
GLUE~\cite{glue} in the second category. Future
experiments in the first category include NT200+ at Lake
Baikal~\cite{baikal_limit}, ANTARES in the Mediterranean~\cite{antares},
NESTOR in Greece~\cite{nestor}, AMANDA-II~\cite{amandaII}, and
ICECUBE~\cite{icecube}, the
next-generation version of AMANDA, whereas the air shower based
category includes the Auger project~\cite{auger_nu}, the Japanese
telescope array~\cite{ta_nu}, the fluorescence/\v{C}erenkov detector
MOUNT~\cite{mount}, and the space based OWL~\cite{owl_nu} and
EUSO~\cite{euso} experiments. The vertical bars for the MOUNT
sensitivity characterize the uncertainties due to the not yet
determined zenith angle range and sensitivity to the fluorescence
component. The
OWL sensitivity estimate is based on deeply penetrating atmospheric
showers induced by electron or muon-neutrinos only~\cite{owl_nu}
and may thus be considerably better if tau neutrinos, \v{C}erenkov
events, and Earth skimming events are taken into account~\cite{skim}. The same
applies to the EUSO project~\cite{euso}. The AMANDA-II sensitivity will lie
somewhere between the ANTARES and ICECUBE sensitivities~\cite{amandaII}.
The maximized fluxes shown in Figs.~\ref{F5} and~\ref{F5a} are
considerably higher than the ones discussed in
Refs.~\cite{cosmogenic,stecker,pj,ydjs,ess}, and should be easily detectable
by at least some of these future instruments, as is obvious from
Fig.~\ref{F5a}.

\section{Neutrino Fluxes in Top-Down Scenarios}
\begin{figure}[ht]
\begin{center}
\includegraphics[angle=270,width=.48\textwidth,clip=true]{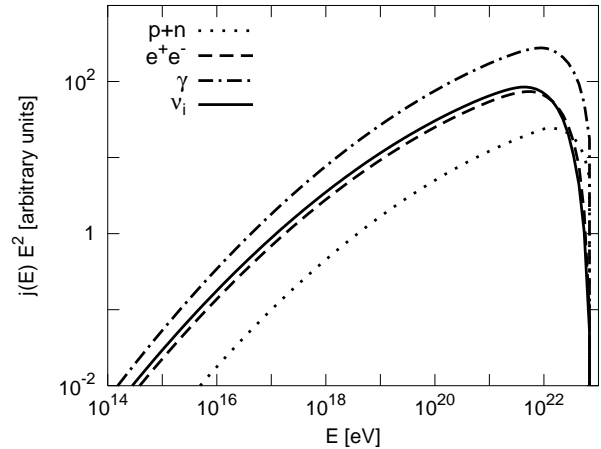}
\end{center}
\caption[...]{Unnormalized nucleon, electron, $\gamma-$ray, and
neutrino (per flavor) MLLA spectra resulting from X particles decaying
into two quarks without supersymmetry. These spectra are used as
injection spectra of top-down models in the present work.
The spectra of antiparticles can be assumed identical.
\label{F7}}
\end{figure}

Historically, top-down scenarios were proposed as an alternative
to acceleration scenarios to explain the huge energies up to
$3\times10^{20}\,$eV observed in the cosmic ray spectrum~\cite{bhs}.
In these top-down scenarios UHECRs are the
decay products of some super-massive ``X'' particles of mass
$m_X\gg10^{20}\,$eV close to the grand unified scale, and have
energies all the way up to $\sim m_X$. Thus,
The massive X particles could be metastable relics of the early Universe
with lifetimes of the order the current age of the Universe
or could be released from topological defects
that were produced in the early Universe during
symmetry-breaking phase transitions envisaged in Grand Unified
Theories (GUTs). The X particles typically decay into leptons
and quarks. The quarks hadronize, producing jets of hadrons
which, together with the decay products of the unstable leptons,
result in a large cascade of energetic photons, neutrinos and light
leptons with a small fraction of protons and neutrons, some
of which contribute to the observed UHECR flux. The
resulting injection spectra have been calculated from QCD in
various approximations, see Ref.~\cite{bs} for a review and
Ref.~\cite{frag} for more recent work. In the present work
we will use the spectra shown in Fig.~\ref{F7} which are obtained from
solving the DGLAP equations in modified leading logarithmic
approximation (MLLA) without supersymmetry for X particles
decaying into two quarks, assuming 10\% nucleons in the
fragmentation spectrum.

For dimensional reasons the spatially averaged X particle
injection rate can only
depend on the mass scale $m_X$ and on cosmic time $t$ in the
combination
\begin{equation}
  \dot n_X(t)=\kappa m_X^p t^{-4+p}\,,\label{dotnx}
\end{equation}
where $\kappa$ and $p$ are dimensionless constants whose
value depend on the specific top-down scenario~\cite{bhs}.
Extragalactic topological defect sources usually predict
$p=1$, whereas decaying super-heavy dark matter of lifetime
much larger than the age of the Universe corresponds to
$p=2$~\cite{bs}. In the latter case the observable flux will
be dominated by decaying particles in the galactic halo and
thus at distances smaller than all relevant interaction
lengths. Composition and spectra will thus be directly given
by the injection spectra which are most likely inconsistent
with upper limits on the UHE photon fraction
above $10^{19}\,$eV~\cite{gamma_comp},
see Fig.~\ref{F7}. In addition, decaying dark matter scenarios
suffer in general from a more severe fine tuning problem
and predict a smaller neutrino flux than extragalactic topological
defect model scenarios. We will therefore here focus on the latter,
with $p=1$.

\begin{figure}[ht]
\begin{center}
\includegraphics[angle=270,width=.5\textwidth,clip=true]{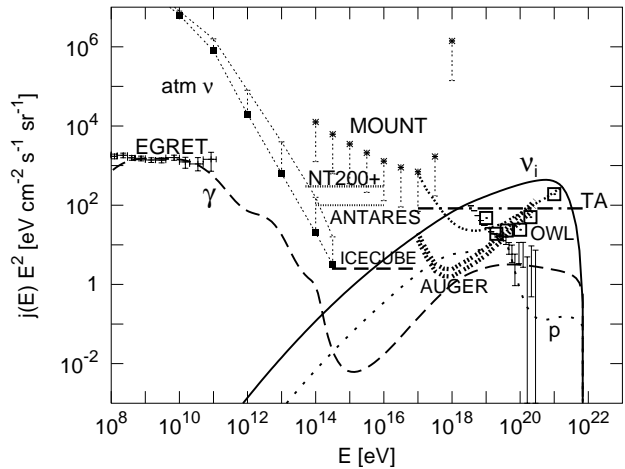}
\end{center}
\caption[...]{Flux predictions for a TD model characterized by $p=1$,
$m_X=2\times10^{14}\,$GeV, and the injection spectra given in Fig.~\ref{F7}.
The line key is as in Fig.~\ref{F5a}.
\label{F8}}
\end{figure}

Fig.~\ref{F8} shows the results for $m_X=2\times10^{14}\,$GeV,
with $B=10^{-12}\,$G, and again the minimal URB consistent with
data~\cite{Clark,pb}. These parameters lead to optimistic neutrino fluxes
for the maximal normalization consistent with all data. For
more detailed recent discussions of top-down fluxes see Refs.~\cite{slsc,slby}.

\section{The Z-Burst Scenario with Acceleration Sources}
\begin{figure}[ht]
\begin{center}
\includegraphics[angle=270,width=.5\textwidth,clip=true]{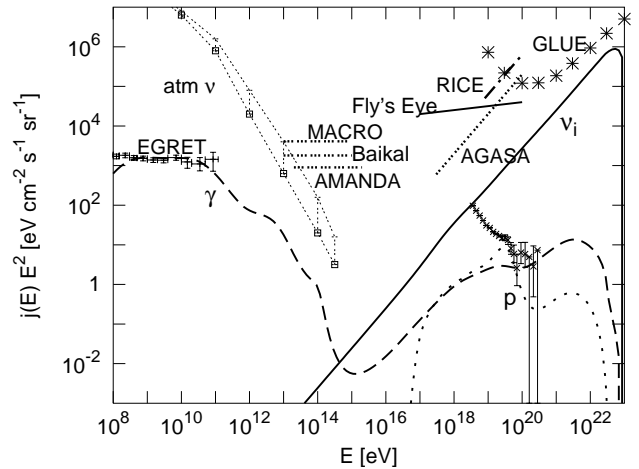}
\end{center}
\caption[...]{Flux predictions for a Z-burst model averaged over flavors and
characterized by the injection parameters $z_{\rm min}=0$, $z_{\rm max}=3$,
$\alpha=1$, $m=0$, in Eq.~(\ref{para_inj}) for neutrino primaries. All
neutrino masses were assumed equal with $m_\nu=0.1$~eV and we again
assumed maximal mixing between all flavors. The line key is as
in Fig.~\ref{F5}.
\label{F9}}
\end{figure}

In the Z-burst scenario UHECRs are produced by Z-bosons
decaying within the distance relevant for the GZK effect. These
Z-bosons are in turn produced by UHE neutrinos interacting with
the relic neutrino background~\cite{zburst1}. If the relic neutrinos
have a mass $m_\nu$, Z-bosons can be resonantly produced by UHE
neutrinos of energy
$E_\nu\simeq M_Z^2/(2m_\nu)\simeq4.2\times10^{21}\,{\rm eV}\,({\rm eV}/m_\nu)$.
The required neutrino
beams could be produced as secondaries of protons accelerated
in high-redshift sources. The fluxes predicted in these scenarios
have recently been discussed in detail in Refs.~\cite{fkr,kkss}.
In Fig.~\ref{F9} we show an optimistic example taken from Ref.~\cite{kkss}
for comparison with the other scenarios. As in Refs.~\cite{fkr,kkss}
no local neutrino over-density was assumed. The sources are
are assumed to not emit any $\gamma-$rays, otherwise the Z-burst
model with acceleration scenarios is ruled out, as demonstrated
in Ref.~\cite{kkss}. We note that no known
astrophysical accelerator exists that meets the requirements
of the Z-burst model~\cite{kkss}. Also note that even exclusively
emitting neutrino sources appear close to be ruled out already
by the GLUE bound.

\section{Mono-energetic Super-heavy Relic Neutrino Sources}
\begin{figure}[ht]
\begin{center}
\includegraphics[angle=270,width=.5\textwidth,clip=true]{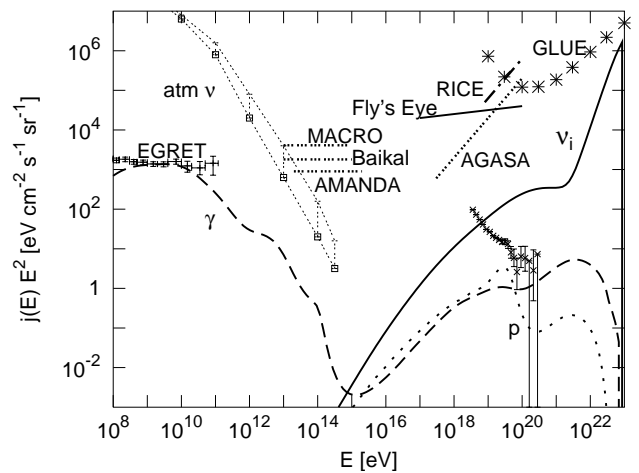}
\end{center}
\caption[...]{Flux predictions for a Gelmini-Kusenko model
characterized by $p=2$, $m_X=2\times10^{14}\,$GeV in Eq.~(\ref{dotnx}),
with X particles exclusively
decaying into neutrino-anti-neutrino pairs of all flavors
(with equal branching ratio), assuming all neutrino masses
$m_\nu=0.1\,$eV. The line key is as in Fig.~\ref{F5}.
\label{F10}}
\end{figure}

Since no known astrophysical accelerator exists that produces sufficiently
strong neutrino beams up to sufficiently high energies for the
Z-burst scenario to work~\cite{kkss}, one can speculate about more
exotic sources. Gelmini and Kusenko~\cite{gk} have considered a
top-down type source for the Z-burst scenario,
namely super-heavy particles mono-energetically decaying exclusively
into neutrinos. In Fig.~\ref{F10} we show predictions for one
example of this kind of model with the same maximal energy
$E_{\rm max}=10^{23}~{\rm eV} = m_X/2$  as for case of Z-burst model.
Again, no local neutrino over-density was assumed and the calculation
assumed a minimal URB flux and a magnetic field $B=10^{-12}\,$G.
Note that, for the same UHECR flux, in this scenario the photon
flux in the EGRET region is larger than in the Z-burst model with
power-law neutrino sources, compare Fig.~\ref{F9}, because all secondary
photons from Z-boson decays at redshifts $z>3$ contribute only to the
EGRET energy range. Thus, the normalization of the photon flux
to the EGRET measurements leads to a decrease of the UHE proton and
photon fluxes, see Fig.~\ref{F10}. The EGRET measurement therefore
considerably constrains the parameter space of this model.
Neutrino masses $m_\nu \gg 0.1$ eV are required, which allow
X-particle masses $m_X\lesssim10^{14}\,$GeV implying
secondary photon fluxes that are below the measured level in the
EGRET energy range.

Let us also note though the neutrino flux in UHECR region is much smaller 
in Gelmini-Kusenko model in compare with Z-burst model, the GLUE bound 
constrains both models in the peak of neutrino flux in similar way.

\section{Neutrino fluxes in a non-shock acceleration model for AGN}

\begin{figure}[ht]
\begin{center}
\includegraphics[angle=270,width=.5\textwidth,clip=true]{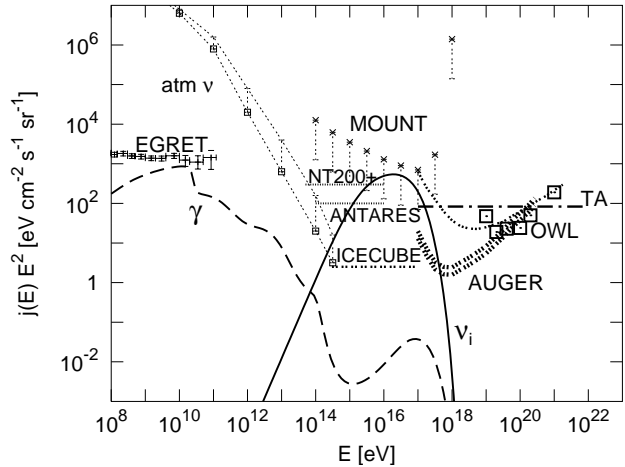}
\end{center}
\caption[...]{Neutrino flux predictions for the AGN model~\cite{neronov}
for a uniform distribution of blazars (no redshift evolution).
The photon flux is below measured EGRET value. The typical neutrino
flux in this model contains the same energy as the photons.
The position of the peak is governed by the initial proton distribution.
The line key is as in Fig.~\ref{F5a}.
\label{F12}}
\end{figure}

Although the recent exciting discoveries by the Chandra
X-ray space observatory added much to our knowledge of structures of
the jets of powerful radiogalaxies
and quasars, they didn't solve the old problems and, in fact,
brought new puzzles. If the observed X-ray emission is due to
synchrotron energy losses of electrons,
the energy of such electrons should be of the order of 100 TeV. Electrons
of such energies loose all their energy on a typical scale of only 0.1 kpc.
In order to explain observed X-ray data, such  100 TeV electrons
should be created uniformly over the jet length of order of 100 kpc.
The conventional shock-wave
acceleration mechanism in this case would require a jet uniformly filled with
1000 identical shocks, following one another along the jet axis.

In Ref.~\cite{neronov} a ``non-shock acceleration''  version
of the electron synchrotron model was proposed, namely assuming that
electrons are not accelerated
in the jet, but are instead the result of pair production by
very high energy (VHE) $\gamma-$rays interacting with the
CMB.  The typical attenuation length of
$10^{16-18}$ eV photons is of order 100 kpc, comparable with the lengths of
large scale extragalactic jets.

In this model the  VHE $\gamma$-rays are produced by accelerated protons
interacting with the ambient photon fields
(supplied, for example  by the accretion disk around the massive black hole)
through photo-meson processes. At the same time those protons produce neutrinos
which are emitted in the direction of the jet. Therefore, this model
predicts a high neutrino flux comparable in power with the 
$\gamma$-ray flux. The detailed numerical simulations of proton acceleration
in the central engine of the AGN~\cite{new} show that
the collimated jet of almost mono-energetic VHE protons (linear accelerator)
can be created in the electro-magnetic field around the black hole
and the energy of those protons can be converted into photons and neutrinos,
while protons can be captured inside of the source.
The nucleon flux leaving the AGN is well below observed cosmic ray
flux in this scenario. Furthermore, since all nucleons leaving the
source are well below the GZK cutoff, there is no cosmogenic contribution
to the neutrino flux from these sources.

Fig.~\ref{F12} shows a typical prediction of the diffuse neutrino
flux in this scenario. This flux is beyond the WB limit which is not
applicable in this case because the sources are optically thick for
nucleons with respect to pion photo-production.
The flux is consistent with MPR bound for 
optically thick sources.

In the AGN  model discussed above, blazars would be seen by neutrino
telescopes as point-like sources with neutrino fluxes which are smaller or
of the same order as the photon flux emitted by these same sources and
which are detectable by $\gamma$-ray telescopes.

\section{Conclusions}
Based on our transport code we reconsidered neutrino flux predictions
and especially their maxima consistent with all cosmic and
$\gamma-$ray data, for cosmogenic neutrinos produced through
pion production of UHECRs during propagation, and for the
more speculative Z-burst scenario and top-down scenarios.
We pointed out that one can easily exceed the WB bound and, in the
most optimistic cases, even the MPR bound for cosmogenic
neutrinos in scenarios with cosmic ray injection spectra harder than
$E^{-1.5}$, maximal energies $E_{\rm max}\gtrsim10^{22}\,$eV,
and redshift evolution typical for quasars, or stronger. Given our poor
knowledge on the origin of UHECRs, in our opinion these are possibilities
that should not be discarded at present, especially since they
would lead to considerably increased prospects of ultra-high energy
neutrino detection in the near future.
We also show that for non-shock AGN acceleration models
the AGN neutrino fluxes can reach the $\gamma-$ray bound Eq.~(\ref{egret})
around $10^{16}\,$eV which represents the ultimate limit for all
scenarios of $\gamma-$ray and neutrino production involving
pion production.

Finally, we note that fluxes as high as for the optimistic scenarios
discussed here would also lead to much stronger constraints
on the neutrino-nucleon cross section at energies beyond the electroweak
scale. This is important, for example, in the context of theories
with extra dimensions and a fundamental gravity scale in the TeV
range~\cite{tev}.

\section*{Acknowledgments}
We would like to thank Andrey Neronov and Igor Tkachev for
fruitful discussions and comments. We thank Cyrille Barbot,
Zoltan Fodor, and Andreas Ringwald for useful comments on
the first version of the manuscript. We also thank David Seckel
for information on the RICE experiment.

%%%%%%%%%%%%%%%%%%%%%%%%%%%%%%%%%%%%%%%%%%%%%%%%%%%%%%%
%%%%%%%%%%%%%%%%%%%%%%%%%%%%%%%%%%%%%%%%%%%%%%%%%%%%%%%%%%%%%%%%%%

\end{document}